\documentclass[traditabstract]{aa} 
\usepackage{graphicx}
\usepackage{txfonts}

\begin{document}

   \title{PF191012 Myszyniec - highest Orionid meteor ever recorded}

   \author{A. Olech\inst{1}, P. \.Zo{\l}\c{a}dek\inst{2}, M. Wi\'sniewski\inst{2,3},
K. Fietkiewicz\inst{2}, M. Maciejewski\inst{2}, Z. Tymi\'nski\inst{4}, 
T. Krzy\.zanowski\inst{2}, M. Krasnowski\inst{2},
M. Kwinta\inst{2}, M. Myszkiewicz\inst{2}, K. Polakowski\inst{2},
          \and
          P. Zar\c{e}ba\inst{2}
          }

   \institute{Copernicus Astronomical Center, Polish Academy of Sciences,
              ul. Bartycka 18, 00-716 Warszawa, Poland
	\and
	Comets and Meteors Workshop, ul. Bartycka 18, 00-716 Warszawa, Poland
	\and
Central Office of Measures, ul. Elektoralna 2, 00-139 Warsaw, Poland
	\and
Narodowe Centrum Bada\'n J\c{a}drowych, O\'srodek Radioizotop\'ow POLATOM, 
ul. So{\l}tana 7, 05-400 Otwock, Poland
             }

   \date{Received April 18, 2013; accepted ................., 2013}

\abstract{

On the night of Oct 18/19, 2012 at 00:23 UT a $-14.7$ mag
Orionid fireball occurred over northeastern Poland. The precise
orbit and atmospheric trajectory of the event is presented, based on
the data collected by five video and one photographic {\sl Polish
Fireball Network (PFN)} stations. The beginning height of the meteor
is $168.4\pm0.6$ km which makes the PF191012 Myszyniec fireball the
highest ever observed, well documented meteor not belonging to the
Leonid shower. The ablation became the dominant source of light of the meteor 
at a height of around 115 km. The thermalization of sputtered particles
is suggested to be the source of radiation above that value. The
transition height of 115 km is 10-15 km below the transition heights
derived for Leonids and it might suggest that the material of Leonids
should be more fragile and have probably smaller bulk density than in
case of Orionids.

}


\authorrunning{A. Olech, P. {\.Z}o{\l}\c{a}dek, M. Wi\'sniewski et al.}
\titlerunning{PF191013 Myszyniec fireball}
\maketitle

\section{Introduction}

For many years it was commonly accepted that meteor ablation starts at
a height of around 130 km. It was justified by the trajectories obtained
from photographic and video double station observations which indicated
that the vast majority of meteor events emit their light at heights between
70 and 120 km (Ceplecha 1968).

The television and photographic observations of the Leonid shower in
1995 and 1996 carried out by Fujiwara et al. (1998) showed that the
fastest meteors could start at heights of 130-160 km. It was quickly
confirmed by Spurn\'y et al. (2000a) who reported several 1998 Leonid
fireballs with beginning heights at 150-200 km. Spurn\'y et al. (2000b)
analyzed the radiation type of highest Leonid meteors and suggested that
light emitted over 130 km might be due to the processes not connected
with ablation. All high-altitude meteors from their sample showed
comet-like diffuse structures above 130 km which evolved into typical
moving droplets at normal heights. Spurn\'y et al. (2000b) divided
the light curves of high-altitude meteors into three distinct phases:
diffuse, intermediate and sharp. The sharp phase was connected with
well known ablation process. The light emitted during the diffuse phase
cannot be explained by standard ablation theory and a new type of
radiation has to be taken into account.

The source of the meteor radiation above 130 km was not recognized
until mid 2000s. At that time Hill et al. (2004), Popova et al. (2007)
and Vinkovi\'{c} (2007) suggested that thermalization of sputtered
particles can be the source of diffuse radiation from high altitude
meteors. Both model light curves and theoretical shapes of the moving
bodies were in very good agreement with observations.

Up to that date the extreme heights of meteors were recorded only for
Leonid fireballs. It was not surprising due to the fact that Leonids are
characterized by one of the highest entry velocity among all meteor
showers. However, Koten et al. (2001, 2006) found that high-altitude
meteors could be found among sporadic meteors and Lyrid, Perseid and
$\eta$-Aquariid showers. Still, among the highest meteors (with
beginning height $h_b>160$ km), almost all events were recognized
as Leonids. There was no single high-altitude Orionid meteor in that
sample, which was surprising due to the fact that Orionids are
characterized with high geocentric velocity.

\begin{figure}
 \vspace{14cm}
 \caption{Images of the PF191012 Myszyniec fireball recorded at Siedlce PFN43 (upper
panel) and B{\l}onie PFN42 (lower panel) stations.}
\includegraphics{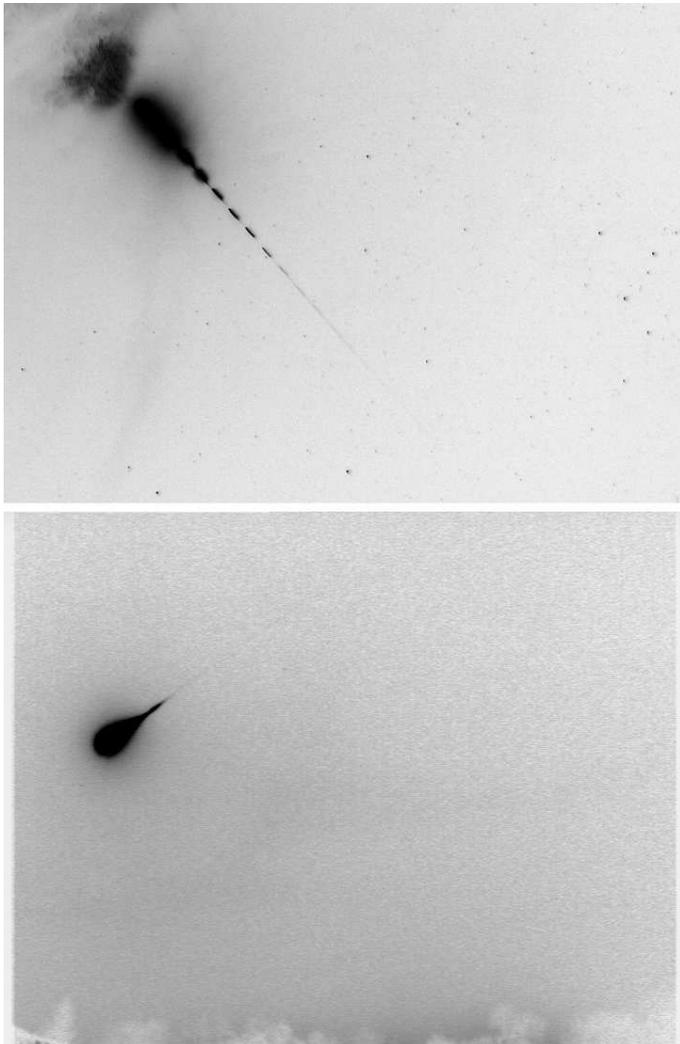}
\end{figure}

\begin{table*}[!t]
\caption[]{Basic data on the PFN stations which recorded PF191012 Myszyniec fireball.}
\centering
\begin{tabular}{|l|l|c|c|c|l|l|l|}
\hline
\hline
Code & Site & Longitude [$^\circ$] & Latitude [$^\circ$] & Elev. [m] & Camera & Lens & Remarks \\
\hline
PFN42 & B{\l}onie & 20.6223 E & 52.1888 N & 86 & Tayama C3102-01A1 & Computar 4 mm f/1.2 & flash saturated \\
PFN32 & Che{\l}m  & 23.4982 E & 51.1355 N & 194 & Mintron MTV-12V8HC & Computar 3.8 mm f/0.8 & end not detected\\
PFN06 & Krak\'ow  & 19.9425 E & 50.0216 N & 250 & Mintron MTV-23X11C & Ernitec 4 mm f/1.2 & low altitude \\
PFN38 & Podg\'orzyn & 15.6817 E & 50.8328 N & 369 & Tayama C3102-01A1 & Computar 4 mm f/1.2 & low altitude \\
PFN43 & Siedlce & 22.2833 E & 52.2015 N & 152 & Canon EOS 350D & Samyang 8 mm f/3.5 & photo with shutter\\
PFN20 & Urz\c{e}d\'ow & 22.1456 E & 50.9947 N & 210 & Tayama C3102-01A1 & Ernitec 4 mm f/1.2 & short path\\
\hline
\hline
\end{tabular}
\end{table*}

\begin{table*}[!t]
\caption[]{Orbital elements of the PF191012 fireball compared to the mean orbit of Orionid 
shower and orbit of comet 1P/Halley. All values are in J2000.0 equinox.}
\centering
\begin{tabular}{|l|c|c|c|c|c|c|c|c|c|c|}
\hline
\hline
  & $\alpha_G$ & $\delta_G$ & $V_G$ & $1/a$ & $e$ & $q$ & $\omega$ & $\Omega$ & $i$ & $D'$ \\
  & [deg]      & [deg] & [km/s] & [1/AU] & & [AU] & [deg] & [deg] & [deg] & \\ 
\hline
PF191012 & 93.00(6) & 14.89(14) & 66.9(8) & 0.0652(69) & 0.961(40) & 0.605(13) & 79(3) & 25.85089(1) & 162.3(3) & \\
Orionids  & 92.5 & 15.7 & 66.8 & 0.0689 & 0.961 & 0.576 &  81.90 & 27.70 & 164.00 & 0.0283\\
1P/Halley &  -   & -    & 65.9 & 0.0550 & 0.967 & 0.582 & 111.80 & 58.10 & 162.26 & 0.0587\\
\hline
\hline
\end{tabular}
\end{table*}  

In this paper we report a detection of the PF191012 Myszyniec fireball
belonging to the Orionid shower, which is one of the highest meteors
ever observed.

\section{Observations}

The {\sl Polish Fireball Network (PFN)} is the project whose main goal
is regularly monitoring the sky over Poland in order to detect
bright fireballs occurring over the whole territory of the country (Olech et
al. 2005, \.Zo{\l}\c{a}dek et al. 2007, 2009, Wi\'sniewski et al. 2012).
It is kept by amateur astronomers associated in {\sl Comets and Meteors
Workshop (CMW)} and coordinated by astronomers from Copernicus
Astronomical Center in Warsaw, Poland. Presently, there are 18 video and
3 photographic fireball stations belonging to {\sl PFN} which operate
during each clear night.

On the night of Oct 18/19, 2012, at 00:23:12 UT, five video and one
photographic stations of the {\sl PFN} recorded bright, $-14.7$ magnitude
fireball belonging to the Orionid shower. Basic properties on the stations
contributing the data to this work are listed in Table 1. Fig. 1 shows
images of the fireball captured by the photographic station in Siedlce and
video station in B{\l}onie.

All video stations contributing to this paper work in PAL
interlaced resolution ($768\times 576$ pixels) with 25 frames per sec 
offering 0.04 sec temporal resolution. The photographic station in
Siedlce works in reduced resolution of $2496\times 1664$ pixels with
30 sec exposure times at ISO 1600. The frequency of rotating shutter
is 10.68 Hz.

\section{Calculations}

The astrometry in the photographic image recorded at PFN43 Siedlce
station was performed using {\sc Astro Record 3.0} software (de Lignie
1997). The accuracy of the meteor path determination reached 10
arcmin. In the case of the video data the modified Turner method was used as
described in Olech et al. (2006). For the video images the accuracy of
the astrometry varied from 9 arcmin for PFN42 B{\l}onie station to
11 arcmin at PFN32 Che{\l}m station. The stations at Krak\'ow,
Podg\'orzyn and Urz\c{e}d\'ow were located too far from the meteor for
using their data in the final calculations.

The trajectory and orbit of the PF191012 Myszyniec fireball was computed
using {\sc PyFN} software written by P. \.Zo{\l}\c{a}dek
(\.Zo{\l}\c{a}dek 2012). {\sc PyFN} is written in Python with usage of
SciPy module and CSPICE library. For trajectory and orbit computation it
uses the plane intersection method described by Ceplecha (1987).

{\sc PyFN} accepts data in both {\sc MetRec} (Molau 1999) and {\sc
UFOAnalyzer} (SonotaCo 2009) formats and allows for semi-automatic
search for double-station meteors. Once the trajectory and orbit is
computed, it can be compared to the other orbits in the database using
Drummond criterion $D'$ (Drummond 1979). Moreover it allows to compute
mean $D'$ value in the vicinity of the meteor, which can be used for
searching for new meteor showers as was demonstrated by
\.Zo{\l}\c{a}dek and Wi\'sniewski (2012).

\section{Results}

Table 2 lists the radiant parameters and orbital elements of the
PF191012 fireball computed from our data and compared to the mean
photographic orbit of Orionid shower and orbit of comet 1P/Halley
(Lindblad and Porubcan 1999). Both the radiant and orbital elements
clearly show that there is no doubt that the PF191012 fireball belongs to
the Orionid shower. It is also confirmed by the value of $D'$ criterion,
which in both cases is significantly smaller than 0.105 indicating that
all discussed bodies are related.

\begin{table}[!t]
\caption[]{The basic trajectory and radiant data of the PF191013 Myszyniec fireball}
\centering  
\begin{tabular}{lccc}
\hline
\multicolumn{4}{c}{2012 Oct 19, ${\rm T} = 00^h23^m12^s \pm 1^s$ UT}\\
\hline
\multicolumn{4}{c}{Atmospheric trajectory data}\\
\hline
 & {\bf Beginning} & {\bf Max. light} & {\bf Terminal} \\
Height [km] & $168.4\pm0.6$ & $77.7\pm1.0$ & $69.4\pm0.6$ \\
Long. [$^\circ$E] & $22.336\pm0.005$ & $21.186\pm0.020$ & $21.040\pm0.002$\\
Lat. [$^\circ$N] & $52.865\pm0.004$ &  $53.385\pm0.010$ & $53.463\pm0.007$\\
Abs. mag & $1.5\pm1.0$ & $-14.7\pm1.0$ & $-1.0\pm0.5$ \\
Slope [$^\circ$] & $42.4\pm0.5$ & $41.5\pm0.5$ & $41.4\pm0.6$ \\
Duration [s] & \multicolumn{3}{c}{$2.19\pm0.04$}\\
Length [km] & \multicolumn{3}{c}{$148.4\pm0.8$}\\
Stations & \multicolumn{3}{c}{Siedlce, B{\l}onie, Che{\l}m}\\
\hline
\multicolumn{4}{c}{Radiant data (J2000.0)}\\
\hline
 & {\bf Observed} & {\bf Geocentric} & {\bf Heliocentric} \\
RA [$^\circ$] & $92.71\pm0.13$ & $93.00\pm0.06$ & --- \\
Decl. [$^\circ$] & $15.08\pm0.09$ & $14.89\pm0.14$ & --- \\
Vel. [km/s] & $68.0\pm0.7$ & $66.89\pm0.76$ & $41.51\pm0.82$\\
\hline
\end{tabular}
\end{table}

\begin{figure}
 \vspace{7.1cm}
 \caption{The luminous trajectory of the PF191013 fireball over northeastern Poland 
and the location of three PFN stations which data were used in calculations.}
\includegraphics{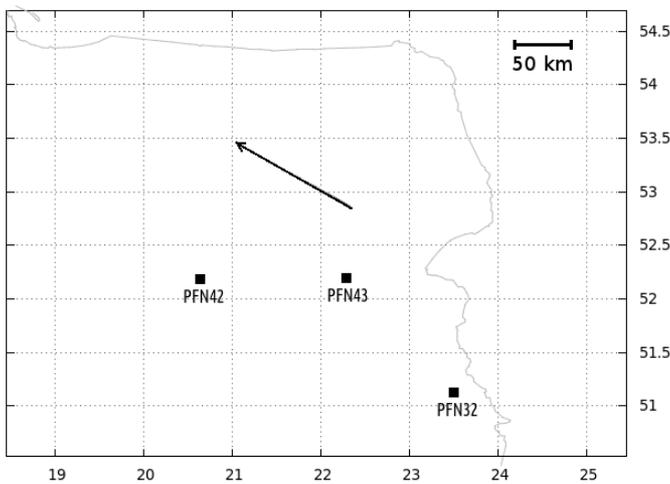}
\end{figure}

The trajectory and radiant parameters derived from the data collected in
three {\sl PFN} stations are summarized in Table 3. Additionally, Fig. 2
shows the map of northeastern Poland with the location of Siedlce,
B{\l}onie and Che{\l}m stations and the luminous trajectory of the
PF191013 fireball.

The meteoroid entered the atmosphere at a height of $168.4\pm0.6$ km
which makes it the highest Orionid meteor ever observed. The initial
velocity and absolute magnitude were $68.0\pm0.7$ km/s and $1.5\pm1.0$,
respectively. These data come from PFN32 Che{\l}m station where the
beginning of the meteor was recorded. The earliest phase of the luminous
path was located outside the field of view of the images from PFN42
B{\l}onie station. The meteor entered this field at height of 
$162.8\pm0.6$ km.

Although some kind of diffuse structure of the meteor shape at high
altitude is visible in both B{\l}onie and Che{\l}m data, it can be
explained by combination of high angular velocity of the bolide and
interlace effect on 0.04 sec temporal resolution data.

The first trace of the path of the meteor in the photo taken by PFN43
Siedlce station is located at height of $152.0\pm0.5$ km. Due to the
lack of shutter breaks on the green path recorded in the heights
range 128-152 km, we conclude that this part of the recorded light
comes not from the meteor itself but from its persistent train.

The first clear shutter break is detected at height of 128 km. At
this point the color of the meteor changes from green to white-yellow.
Below the height of 100 km color again changes and meteor starts to
show white-blue hue.

After initial oscillations, the magnitude of the meteor was monotonically
increasing. At a height of 87 km the first, small flare occurred. The
second and main flare with the maximum absolute magnitude of
$-14.7\pm1.0$ was observed at a height of 77.7 km. The third and final
flare occurred at 74 km. The final flare did not cause the complete
disintegration of the body. Small fragment with brightness of
$-1.0\pm0.5$ mag was observed over the next three video frames and
disappeared at a height of 69 km.

The appearance of the flares is concluded not from the light curve
of the fireball but from the shape of the train recorded in the images
from B{\l}onie station (see upper panel of Fig. 3).

Koten et al. (2006) derived the empirical relations between beginning
height of the meteor and its photometric mass. The relations differed in
slope between all meteors in their sample and high altitude meteors from
Leonid shower. They suggested that for high altitude meteors the steeper
relation should be taken into account. However, all points for high
altitude Perseids, which are slower than Leonids, are located clearly below
the steeper relation. Knowing that Orionids are also slower than Leonids we
decided to use traditional relation given by their equation (1). According
to this, the photometric mass of the PF191013 Myszyniec fireball is
$360\pm110$ grams.

The empirical relations of Koten et al. (2006) are only rough
estimations, thus we decided to calculate the initial mass of the
meteoroid using the radiation equation of meteors (Ceplecha 1996). 
For determination of the luminous efficiency we used ReVelle and
Ceplecha (2001) approximation for fast meteors. The resulting photometric
mass is around 1500 grams. Due to the $\sim 1$ mag error in
brightness determination it is only rough estimate. The real value
of the initial mass should be in 600-3500 grams range.

\begin{figure}
 \vspace{17cm}
 \caption{The persistent train of the PF191013 fireball recorded at
video frames in PFN42 B{\l}onie station (upper image) and six consecutive
30 second exposures obtained in PFN43 Siedlce station.}
\includegraphics{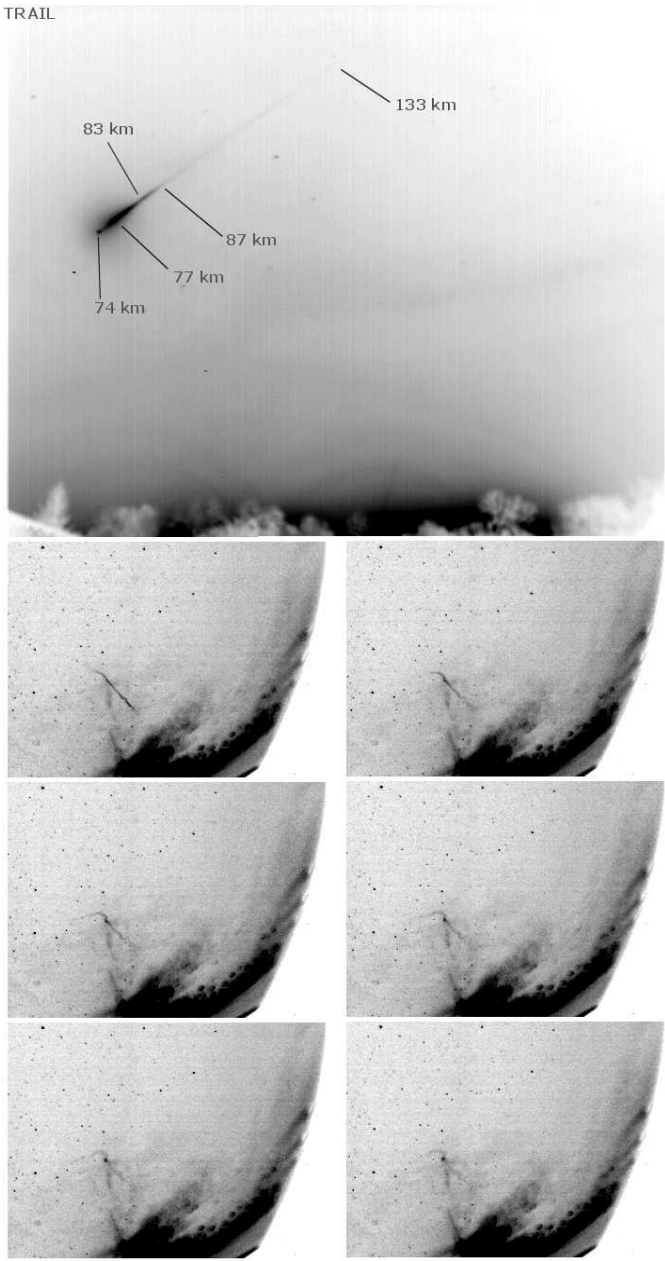}
\end{figure}

The meteor left an extremely bright persistent train. Its peak brightness
reached $-13$ magnitude. During the first phase the brightness of the
train was decreasing at a rate of 2 magnitudes per second. Evolution of
the train was recorded on 38 images (30 seconds exposure each) captured
by the Siedlce station and is shown in Fig. 3. Additionally, the light
curve of the train is presented in Fig. 4.

The evolution of the absolute magnitude of the meteor in the function of
height is presented in Fig. 5. The oscillation of the magnitude of the
meteor at high altitude was already reported by Spurn\'y et al.
(2000a,b) and Koten et al. (2006) and is typical for high-altitude
meteors. Clear change of the slope of magnitude increase is evident at a
height of around 115 km. At this point the ablation process becomes the
dominant source of light.

\begin{figure}
 \vspace{6.8cm}
 \caption{The evolution of the absolute magnitude of the persistent train left by
the PF191012 fireball.}
\includegraphics{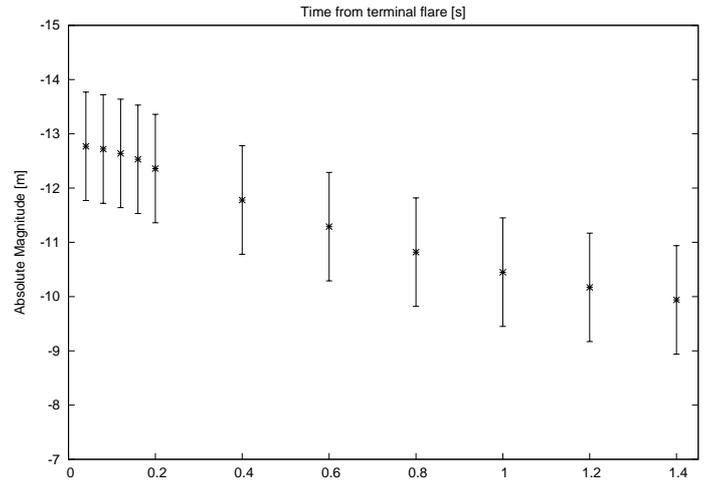}
\end{figure}

\begin{figure}
 \vspace{6.8cm}
 \caption{The evolution of the absolute magnitude of the PF191012 fireball.}
\includegraphics{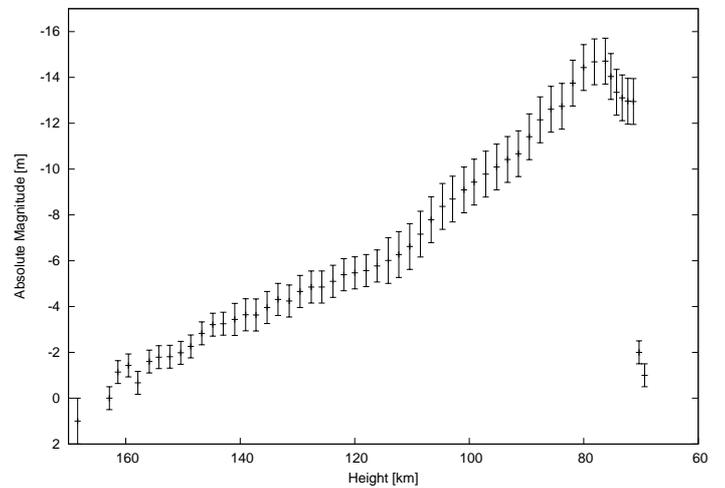}
\end{figure}

Detailed inspection of the light curve shows some differences
between the Orionid fireball light curve and these obtained for faster
Leonids. Vinkovi\'{c} (2007) derived the theoretical light curves for
high altitude meteors for the beginning heights of 200 and 171 km. In
both cases, at high altitude of 145-160 km, one can see slight change of
the slope from the steeper to gentle one, however this change is more
pronounced in the lower meteor. In our case such a change is only barely
visible at height of slightly below 150 km. It can be understood taking
into account the fact that amount of generated light in the sputtering
process depends on the number of collisions and this number is lower for
slower meteors. The light curve of the Orionid fireball should resemble
Leonids light curves obtained for higher meteors.

Additional and more pronounced change of the slope of the brightness of
PF191013 Myszyniec fireball is recorded at height of 115 km. It can be
interpreted as transition from the intermediate phase (where both
sputtering and ablation work together) to final (sharp) phase where only
ablation process is responsible for generating the light. What is
interesting, obtained value of 115 km is 10-15 km below the transition
heights derived for Leonids by Spurn\'y et al. (2000b) and Koten et al.
(2006). It may suggest that the material of Leonids should be more
fragile and have probably smaller bulk density. It is in agreement with
recent results of Borovi\v{c}ka (2007) or Babadzhanov and Kokhirova (2009).

\section{Summary}

In this paper we presented an analysis of the multi-station
observations of a bright fireball belonging the Orionid meteor
shower. Our main conclusions are as follows:

\begin{itemize}

\item the meteor started on 2012 Oct 18/19 at 00:23:12 UT over
the northeastern part of Poland and was detected by five video and one
photographic stations of {\sl Polish Fireball Network},

\item the orbital elements and radiant parameters of the meteor
clearly show that it belongs to the Orionid shower,

\item the fireball started at a height of $168.4\pm0.6$ km which makes
it the highest Orionid meteor ever observed and one of the highest
meteors known,

\item the initial brightness of the meteor was $1.5\pm1.0$ mag and was
oscillating,

\item the ablation became the dominant source of light at a height
of around 115 km, where we have noticed a clear change of slope of
the magnitude increase,

\item the fireball reached its maximum absolute magnitude of
$-14.7\pm1.0$ at height of $77.7\pm1.0$ km,

\item the persistent train left by the fireball had a maximum absolute
magnitude of around $-13$ and was observed over the next 20 minutes both
in video frames and photographic images.

\end{itemize}

\begin{acknowledgements}
This work was supported Siemens Building Technologies found.
\end{acknowledgements}

\end{document}